\documentclass[12pt]{article}
\begin{document}
\vspace*{-.6in}
\thispagestyle{empty}
\begin{flushright}
CALT-68-2188\\
hep-th/9807195
\end{flushright}
\baselineskip = 20pt

\vspace{.5in}
{\Large
\begin{center}
{\bf Beyond Gauge Theories}\,\footnote{Work
supported in part by the U.S. Dept. of Energy under Grant No.
DE-FG03-92-ER40701.}
\end{center}}
\vspace{.5in}
\begin{center}
John H. Schwarz\\
{\em California Institute of Technology}\\
{\em Pasadena, CA  91125, USA}
\end{center}
\vspace{1in}

\begin{center}
\textbf{Abstract}
\end{center}
\begin{quotation}
\noindent  
Superstring theory, and a recent extension called M theory, are leading candidates
for a quantum theory that unifies gravity with the other forces. As such, they are
certainly not ordinary quantum field theories. However, recent duality conjectures
suggest that a more complete definition of these theories can be provided
by the large $N$ limits of suitably chosen $U(N)$ 
gauge theories associated to the
asymptotic boundary of spacetime.
\end{quotation}

\vfil
\centerline{\it Presented at WEIN 98 in Santa Fe, NM (June 1998)}

\newpage

\pagenumbering{arabic}

\section{Introduction}

The conference organizers suggested to me that I speak on the topic ``Beyond
Gauge Theories," which is what I am doing.  No doubt, they anticipated that my
response would be  that beyond gauge theory there is
{\em superstring theory} and {\em M theory}, and (of course)
it is.  However, until recently we didn't have a complete nonperturbative
definition of what these are, even though we have learned much about them over
the years.  So the question that has been gradually coming into focus is ``what
underlies superstring theory and M theory?"  The answer that has emerged
recently is {\em gauge theories}.  This means that certain gauge theories
completely define consistent quantum gravity vacua!  After 30 years, we've come
full circle---but much has been learned in the process.

For those of you who haven't been following this subject, or are new to the
field, let me recall that string theory was developed in the period 1968--73 as
a phenomenological theory of the strong interactions.\footnote{For a collection of 
review articles from this period  see \cite{jacob} and \cite{scherk75}.}
String eigenmodes were
identified as {\em hadrons} and rules for computing scattering amplitudes were
developed.  This program had some qualitative successes, such as incorporating
{\em Regge behavior} and {\em duality}.  There were serious problems, however,
and the string approach to describing strong interactions was dropped (by
almost everybody) in the mid-70's for two main reasons:  First, consistent
string theories had numerous unrealistic features such as ten dimensions,
massless particles (with $J\leq 2$), supersymmetry, no parton-like structures, etc.
Second, QCD was developed and quickly recognized to be a compelling
alternative.

In 1974, Jo\"el Scherk and I proposed that since string theory incorporates
general relativity~\cite{yoneya74} as well as gauge theory and is ultraviolet finite, it should
be considered a candidate for a unified quantum theory of gravity and all other
forces~\cite{scherk74}.  This meant that the characteristic string length scale $\ell_s$ should
be close to the Planck scale $(10^{-32} cm)$ rather than the QCD scale
$(10^{-13} cm)$.  Equivalently, the string mass scale $m_s$ should be about
$10^{18}$ GeV rather than 100 MeV.

Before giving you a progress report on the quarter-century-old quest 
of constructing a unified quantum theory
of gravity and all other forces in terms of a unique underlying 
superstring theory that has no 
arbitrary dimensionless parameters, I would like to make a short digression.
When we talk about this subject to our experimental friends, quite understandably
they want to know whether our theory can be regarded as an extension of the standard model,
and if so, what experimentally testable predictions does it make. In fact, in the
13+ years since the pioneering paper of 
Candelas {\it et al.}~\cite{candelas85}, 
there have been many attempts to construct realistic superstring vacua.
The conclusion of these studies is that it is possible to find ones that have many qualitatively
correct features, but (at the present time)
we cannot single out a particular one of them as a compelling
candidate.  The only general feature that they all seem to share is low-energy
({\it i.e.}, electroweak scale) supersymmetry. 
So we are inclined to call {\em supersymmetry} a generic
prediction.  Suppose that the LHC rules this out.  Will we still believe in this
approach? I can only speak for myself, though I suspect that most others working
in this field would agree. I believe that we have found the unique 
mathematical structure that consistently combines quantum mechanics
and general relativity. So it must almost certainly be correct. For this reason,
even though I do expect supersymmetry to be found, I would not abandon this
theory if supersymmetry turns out to be absent.  It is (remotely) conceivable that
all supersymmetries are broken at some very high unification scale, and the gauge hierarchy
problem is solved in some other way. Then the unification of the three gauge couplings
in the minimal supersymmetric extension of the standard model would be just a bizarre
coincidence.

Having pinned me down, you see
that we have no absolute predictions at this point, except that we are working on
the right theory. (We do have some solid {\em
postdictions}, however: general relativity and gauge theory.) The problem is not
our unwillingness to do phenomenology, but rather the strongly held belief
that we need a better understanding of the mathematical structure that we
are dealing with, before we can make reliable predictions. We are still discerning
what the theory is, and in this I will report some significant progress. But even when
that is settled, we will need to understand better how a realistic vacuum is chosen.
In particular, it should have a vanishing cosmological constant (or at least one that is
extremely small). This is easy when supersymmetry is unbroken, but when it is broken
one tends to get a cosmological constant 
that is much too large.\footnote{Superstring theory is the only theory for which this
is a problem, because it is the only quantum theory in which the cosmological constant
can be computed.}
At the present time, no one has proposed
a realistic superstring vacuum with a sufficiently small cosmological constant. There
is no freedom to fine-tune it away, though that is not what we want to do, anyway. The point
of this rather lengthy digression has been to explain why we continue to do abstruse
mathematics without making any predictions. It's not because we are perverse,
or physicists gone astray. It's what we believe is required to achieve our common goal.

\section{The Superstring Revolutions}

In the {\em first superstring revolution} (1984--85) it was established that
there are five consistent string theories, each of which requires ten dimensions
(nine space and one time) and supersymmetry. (For a pedagogical review
see~\cite{green87}.)
For four of them (the two {\em
heterotic} string theories and the two {\em type II superstring} theories) the
fundamental strings are oriented and unbreakable.  For one of them (the {\em
type I superstring} with gauge group SO(32)) the fundamental string is
unoriented and breakable.  The string coupling constant, $g_s$, is a
dimensionless parameter determined as the value of a scalar field (the
dilaton).  It can be used as an expansion parameter in perturbation theory, in
the usual fashion.  Thus a scattering amplitude involving $N$ on-shell
particles with 10-momenta $p_1^\mu, p_2^\mu, \ldots, p_N^\mu$ has an expansion
\begin{equation}\label{1}
T(g_{s}; p_1, \ldots, p_N) = \sum_{n=0}^{\infty} (g_s)^{2n} T_n (p_1, \ldots,
p_N).
\end{equation}
As in quantum field theory, this is an asymptotic expansion.  For the heterotic
and type II theories there is a single $n$-loop Feynman diagram---a closed,
orientable genus-$n$ Riemann surface with $N$ marked points.  (Such a surface
can be visualized as a sphere with $n$ handles.)  Then the $n$-loop amplitude
$T_n$ is given by a $(6n + 2N-6)$-dimensional integral, which has no
ultraviolet divergences.  In the case of the type I theory the expansion is
more complicated.  In the subsequent decade (1985--94), much was learned about string
theory, especially the possibilities for compactifying the extra dimensions.
However, all these studies were within the context of perturbation theory.

The {\em second superstring revolution}, characterized by the discovery of
nonperturbative properties of string theory, began around 1994.  One fact that
quickly became clear is that there is actually a {\em unique} underlying
theory,  with no arbitrary dimensionless parameters~\cite{hull94,witten95a,horava95}. 
(For a review, see \cite{sen98}.) However,
this theory admits many consistent solutions (or quantum vacua),
so it does not uniquely determine the universe we observe.
The five superstring theories that had been identified previously actually
correspond to five limiting points in a very large continuous {\em moduli
space} of consistent quantum vacua.  Moreover, this moduli space has a sixth
limiting point of high symmetry---namely, one with a flat 11-dimensional
spacetime~\cite{townsend95a,witten95a}.  
The quantum theory with this vacuum is called {\em M theory}.  At
low energies, the leading approximation to M theory is 11-dimensional
supergravity.  This is a classical field theory that was discovered 20 years
ago~\cite{cremmer78a}.  By itself, it does not define a consistent quantum theory, but the belief
is that M theory is a well-defined quantum theory.  The challenge, of course,
is to describe it precisely.

The key to recognizing that there is a unique underlying theory was the
discovery of various {\em dualities}, which are labelled by the 
letters  S, T and U.  T dualities relate
large compactification spaces to small ones~\cite{giveon94}.  
For example, a circle of radius
$R$ can be equivalent to one of radius $\ell_s^2/R$.  In this way one relates
the two type II theories and the two heterotic theories.  S duality relates
weak coupling and strong coupling.  For example, one of the heterotic string
theories with coupling constant $g_s$ is equivalent to the type I theory with
coupling constant $g'_s = 1/g_s$.  U dualities combine these notions by also
relating compactification sizes to coupling strengths.

The eleven-dimensional M theory vacuum corresponds to the strong coupling limit
of type IIA superstring theory.  Let me sketch how this works.  The claim is
that the IIA theory, with string coupling $g_s$ and string mass scale $m_s$,
actually has a circular 11th dimension that is invisible in perturbation
theory.  To see how this can happen,  consider an eleven-dimensional theory with mass
scale $m_p$ (the 11d Planck mass) and a circular spatial dimension of radius
$R$.  Then, it turns out that the proper identifications are provided by
\begin{equation}\label{2}
m_s^2 = 2\pi Rm_p^3,
\end{equation}
and
\begin{equation}\label{3}
g_s = 2\pi Rm_s.
\end{equation}
The significance of these formulas will be explained shortly.  First, let us
note that the perturbation expansion of the type IIA theory is an expansion in
powers of $g_s$ for fixed $m_s$.  The second relation shows that this is
equivalent to an expansion about $R = 0$, which shows that the appearance of an
eleventh dimension is nonperturbative.  Decompactification $(R \rightarrow
\infty)$ is achieved in the strong coupling limit ($g_s \rightarrow\infty$).

A second crucial ingredient (the dualities are the first one) in understanding
nonperturbative string theory is the identification and understanding of
various dynamical objects called {\em $p$-branes}~\cite{duff95}.  These are objects whose
energy is concentrated on a spatial surface of $p$ dimensions.  The energy
density, which is usually a constant, is called the {\em tension} of the
$p$-brane.  In this nomenclature, point particles are 0-branes (tension is just
mass in this case), and fundamental strings are 1-branes.  
These objects can be studied with
good mathematical control when they preserve some of the supersymmetry of the
ambient theory.

In most cases of interest\footnote{Type I strings are an exception, which is
why they are breakable.}, the $p$-branes are sources for generalized gauge
fields $A_{\mu_{1}\mu_{2} \ldots \mu_{n}}$.  These fields are antisymmetric in
their Lorentz indices, and can be regarded as generalizations of the Maxwell
field, which has $n = 1$.  An $n$-index gauge field can have an {\em electric}
source with $p = n - 1$ dimensions
or a {\em magnetic} source with $p = D - n - 3$ dimensions.  In
the case of Maxwell theory ($n = 1$,  $D = 4$), electric and magnetic sources both
have $p = 0$ (point particles).  M theory has a 3-index potential, and
therefore in this case $n = 3$ and $D = 11$.  The electric source,
which has $p = 2$, is called the {\em M2-brane},  and the magnetic
source with $p = 5$ is called the {\em M5-brane}.  Type IIB superstring theory
contains various gauge fields, including one with four indices.  Therefore in
this case $n = 4$ and $D = 10$.  The corresponding brane has $p = 3$, and it is
simultaneously electric and magnetic (self-dual).  It is called the {\em
D3-brane}.

Some of the $p$-branes that exist in the type IIA theory can be deduced 
from the relation between M theory and the type IIA
theory discussed above.  Specifically, the M5-brane can give type IIA
$p$-branes with either $p = 4$ or $p = 5$ depending on whether or not one of its
dimensions is wrapped around the circular 11th dimension.  These 
particular $p$-branes are called the
NS5-brane and the D4-brane.  Similarly, the M2-brane can give $p = 1$ or $p =
2$.  The $p = 1$ case is the fundamental string.  This means that the type IIA
superstring is actually a wrapped membrane! This fact is consistent with
eq.~(\ref{2}), which can be intepreted as relating the fundamental string
tension $(2\pi m_s^2)$ to the M2-brane tension $(2\pi m_p^3)$.  The $p = 2$
case is called the D2-brane.

The type IIA theory also contains states that correspond to Kaluza--Klein
excitations of the circular 11th dimension.  The momentum on the circle is
quantized $(p_{11} = N/R)$.  The 10d mass is $M_{10} = \sqrt{M_{11}^2 +
p_{11}^2}$, which for an 11d supergraviton ($M_{11} = 0$) gives $M_{10} = N/R$.
 The $N = 1$ mode has mass
\begin{equation}
M = \frac{1}{R} = 2\pi \frac{m_s}{g_s},
\end{equation}
where we have used eq.~(\ref{3}).  This particle is identified as the D0-brane
of the IIA theory.  Note that its mass diverges at weak coupling,
which means that it is a nonperturbative excitation.

A specific proposal for a nonperturbative formulation of eleven-dimensional M
theory, called {\em Matrix Theory},
 was put forward by Banks {\it  et al.} in 1996~\cite{banks96}.  It is a supersymmetric gauge
theory in which the coordinates of $N$ D0-branes are represented by $N \times
N$ matrices.  The idea, roughly, is to go to the infinite momentum frame by
letting $N \rightarrow \infty$ and $R \rightarrow \infty$ at the same time.
An interpretation of the finite $N$ theory has also been proposed~\cite{susskind97}.
This approach has many features in common with the old parton approach to QCD.
However, thanks to supersymmetry, it seems to be somewhat better defined.
Indeed, many checks of this proposal have been made by comparing Matrix Theory
calculations to graviton scattering amplitudes in 11d supergravity.  For a while
there appeared to be some discrepancies, but now these have all been resolved, so
the conjecture seems to be correct.  Even though this represents important
progress, it is not the last word.  Not only is Matrix Theory awkward to use,
but (more fundamentally) it seems to be applicable to only a limited class of
quantum vacua.  In particular, it does not seem able to describe 
realistic vacua in which
all but four dimensions are compactified.

A special class of $p$-branes in type II superstring theories, a few of which
we have already encountered, are called {\em
D-branes}~\cite{polchinski95}.  Here, $D$ stands for {\em Dirichlet}, because these are branes on
which fundamental strings can end, which is represented by Dirichlet boundary
conditions.  This class of $p$-branes has a number of distinctive
properties.  For one thing they couple to gauge fields in the Ramond--Ramond
sector, which means that they can be represented as bispinors.  Secondly, their
tensions are given by
\begin{equation}
T_{D{p}}= 2\pi m_s^{p + 1}/g_s.
\end{equation}
This dependence on the coupling constant is intermediate between that of
fundamental strings (whose tension is independent of $g_s$) and that of
ordinary solitons, such as the NS5-brane, whose tension is proportional to
$(g_s)^{-2}$.

Another important fact about type II $Dp$-branes comes into play
when $N$ of them are parallel
and (nearly) coincident. In this case the 
$(p + 1)$-dimensional world volume of the branes
contains excitations confined to the vicinity of the branes.
Then there is an effective world-volume theory which is approximated
at low energies by
$U(N)$ gauge theory in $p + 1$ dimensions with maximal supersymmetry (16
conserved supercharges).  The $N^2$ gauge fields and their supersymmetry
partners arise as the ground-state excitations of fundamental strings
connecting pairs of D-branes.

Generalizations of this basic construction of 
supersymmetric gauge theories using D-branes turn
out to have some very important applications.  It can be generalized to more
complicated {\em brane configurations} in which other $p$-branes are also
involved and some of the supersymmetry is broken.  One area of application has
been to the study of black holes~\cite{maldacena96}.  
The basic idea is that for small $g_s$ one
can carry out controlled perturbative string theory
calculations of the microphysics including an
enumeration of degrees of freedom.  For large $g_s$, on the other hand, one has
strong gravitational fields, an event horizon, and a black hole that can be
described by general relativity.  However, in cases with supersymmetry one can
argue that the degrees of freedom cannot change in continuing from small $g_s$
to large $g_s$.  Thus one can count states (i.e., compute the statistical
mechanical entropy) for small $g_s$ and compare to the area of the event
horizon for large $g_s$.  One finds precise agreement with the
Bekenstein--Hawking formula for a wide variety of examples.  
Despite this remarkable achievement, a general
understanding of why the microscopic entropy 
of a black hole should be $1/4$ the area
of the event horizon has not yet been achieved.

A second area of application of brane configurations has been to the study of
exact nonperturbative properties of supersymmetric gauge theories~\cite{giveon98}.  
Many deep
results, including the classic discoveries of Seiberg and Witten~\cite{seiberg94}, can be
understood quite simply in this way.  This is also a very active subject.

\section{AdS/CFT Duality}

Let me now turn to the latest development in this field, which goes by the name
of {\em AdS/CFT duality}.  Here, AdS stands for {\em anti de Sitter space} and
CFT stands for {\em conformal field theory}.  This is a new duality, quite
different from all previous ones, which is a very hot topic at the present
time.  AdS/CFT duality was proposed by Maldacena in November 1997~\cite{maldacena97}.
As is usually the case with such developments, 
there were important prior~\cite{klebanov97} and
subsequent~\cite{gubser98} contributions by many others.  
In the remainder of this talk, I will
sketch the basic ideas.

A $p$-brane, or collection of $p$-branes, gives rise to a certain space-time
geometry and gauge field configuration, which can be analyzed using the
appropriate supergravity field equations.  In a number of cases one finds that
the geometry has an event horizon, giving a higher-dimensional analog
of black holes.  In some of these cases the geometry near the horizon is
approximated by $AdS_{p+2} \times S^{D-p-2}$.  This means that the AdS space
has $p + 2$ dimensions and the remainder of the $D$ dimensions form a sphere of $D -
p - 2$ dimensions.  There are three basic examples that have maximal
supersymmetry (32 conserved supercharges).  A stack of D3 branes in type IIB
superstring theory has near horizon geometry $AdS_5 \times S^5$, a stack of
M2-branes in M theory gives $AdS_4 \times S^7$, and a stack of M5-branes in M
theory gives $AdS_7 \times S^4$.  These solutions to type IIB and 11d
supergravity were discovered in the mid 1980's,\footnote{The $AdS_4 \times S^7$
case is introduced in~\cite{freund}, the  $AdS_7 \times S^4$ case in~\cite{petervn}, and the
$AdS_5 \times S^5$ case in~\cite{kim}.}
but were not pursued in the
context of superstring/M theory until recently.

Let me briefly describe some of the salient features of anti de Sitter space.
$AdS_{n+1}$ is a maximally symmetric spacetime with a negative cosmological
constant.\footnote{Nobody claims that this realistic, only that it is instructive to study.}
It can be described as a hypersurface in flat space by the equation
\begin{equation}
u_1^2 + u_2^2 - v_1^2 - v_2^2 - \ldots - v_n^2 = R^2,
\end{equation}
where $R$ is called the AdS radius.  This spacetime has Lorentzian signature
and reduces to Minkowski spacetime in $n + 1$ dimensions in the limit $R
\rightarrow \infty$.  Just as an $n+1$-dimensional 
sphere ($S^{n+1}$) has $SO(n+2)$ symmetry, the
symmetry of this spacetime is $SO(2, n)$, a noncompact version of the rotation
group in $n + 2$ dimensions.  This contracts to the Poincar\'e
group (consisting of the Lorentz group $SO(1,n)$ and translations)
in the $R \rightarrow \infty$ limit.  An
intrinsic description of $AdS_{n+1}$ is given by the metric
\begin{equation} \label{metric}
ds^2 = \frac{R^2}{z^2} (dz^2 + dx^\mu dx_\mu),\quad  z>0,
\end{equation}
where
\begin{equation}
dx^\mu dx_\mu = dx_1^2 + \ldots + dx_{n-1}^2 - dt^2.
\end{equation}
Note that the $z = 0$ boundary of $AdS_{n + 1}$ is an $n$-dimensional Minkowski
spacetime, aside from a divergent factor.  What matters is the {\em conformal}
structure, which is not sensitive to this divergent factor.  Thus, even though the
boundary is infinitely far from any point in the interior, it is useful to take it seriously.

The $SO(2,n)$ isometries of the $(n+1)$-dimensional anti de Sitter space induce
the group of conformal transformations on its $n$-dimensional Minkowski
boundary.  (Strictly speaking, the boundary should be compactified by adding a
point at infinity.)  The conformal group is therefore also $SO(2,n)$.  Let me
illustrate how this works with a couple of examples.  The $SO(1,n-1)$ subgroup
of $SO(2,n)$ given by Lorentz transformations of the $x^\mu$ corresponds to the
Lorentz group of the boundary.  The important point is that these
transformations map $z = 0$ to $z = 0$, so that they are well-defined on the
boundary.  Another example is the isometry $x^\mu \rightarrow \lambda x^\mu, z
\rightarrow \lambda z$ where $\lambda$ is a positive scale factor.  This
clearly leaves the AdS metric in eq.~(\ref{metric}) invariant and preserves the boundary.  Thus the
corresponding conformal transformations of the boundary are scale
transformations $x^\mu \rightarrow \lambda x^\mu$.

The basic idea of AdS/CFT duality is to identify a conformally invariant field
theory (CFT) on the $n$-dimensional boundary with a suitable quantum gravity theory in
the $(n + 1)$-dimensional AdS {\em bulk}.  To be specific, from now on I will
focus on the $AdS_5 \times S^5$ solution of the IIB superstring theory.

As we discussed, the IIB theory contains a four-index field
$A_{\mu\nu\rho\lambda}$ for which the D3-brane is a source.  It has a field
strength $F_{\mu\nu\rho\lambda\sigma}$, which is self-dual (in ten dimensions).
 In the $AdS_5 \times S^5$ solution of the theory, the field $F$ has a
quantized flux on the sphere.  Schematically,
\begin{equation}
\int_{S^{5}} F = N,
\end{equation}
where $N$ is a positive integer.  This integer determines the radius $R$ of the
$AdS_5$ and of the $S^5$, which are the same.  Aside from a constant numerical
factor, one finds that
\begin{equation}
R = (g_s N)^{1/4} \ell_s.
\end{equation}
Thus the curvatures (which are proportional to $R^{-2}$)
are small compared to the string scale for $g_s N \gg 1$ and small
compared to the Planck scale for $N\gg  1$.  The first limit suppresses stringy
corrections to supergravity, whereas the latter suppresses quantum corrections
to classical string theory.

Maldacena's duality conjecture is that type IIB superstring theory on $AdS_5
\times S^5$ with $N$ units of $F$ flux {\em is equivalent to} ${\mathcal N} =
4, $ $D = 3 + 1 $ $U(N)$ Yang--Mills theory with $g_{YM}^2 = g_s$.  For this
conjecture to be plausible, it is a crucial fact the ${\mathcal N} = 4$ super
Yang--Mills theory~\cite{brink} is a CFT with vanishing beta function, a fact that was
proved in the early 1980s~\cite{mandelstam}.  
As should be clear from our presentation, this
conjecture arose from considering the near-horizon geometry of a stack of $N$
D3-branes, in the limit $N \rightarrow \infty$.  This duality---if
true---implies an amazing fact: the 4d gauge theory, for large $N$, is actually
a 10d string theory!  Well, it is not yet ``proved,'' but the evidence is mounting
rapidly.  

Let me briefly mention some of the supporting evidence for AdS/CFT duality.  (I
will only discuss the $AdS_5\times S^5$ example described above, but there are
similar stories for other examples.)  First, the symmetries match:  the two
dual theories have the same symmetry supergroup $SU(2,2|4)$.  This supergroup
incorporates 32 fermonic symmetries and a bosonic subgroup $SU(2,2) \times
SU(4)$.  $SU(2,2)$ is the double cover of $SO(2,4)$ the isometry group of
$AdS_5$ and the conformal symmetry group of the 4d gauge theory.  $SU(4)$ is
the double cover of $SO(6)$, the isometry group of $S^5$, and it is the {\em
R-symmetry} group of an ${\mathcal N} = 4$ gauge theory in four dimensions.

The AdS/CFT duality conjecture has been made more precise in~\cite{gubser98}.  
These papers have proposed a mapping between the bulk
string theory and the boundary gauge theory.  It gives a one-to-one
correspondence between on-shell particles of the bulk theory and
gauge-invariant operators of the boundary theory.  Moreover, correlation
functions of these gauge-invariant operators are related to the response of the
type IIB theory to boundary conditions for the associated fields.  These
correspondences have been partially verified.  For example, there is a perfect
correspondence between particles belonging to {\em short} representations of
the AdS supersymmetry algebra and {\em chiral primary operators} of the gauge
theory.

The large $N$ limit of $SU(N)$ gauge theories for fixed $\lambda = g_{YM}^2 N$ was
studied by `t Hooft in 1974~\cite{thooft74}.  He showed that only Feynman diagrams of planar
topology contribute in this limit.  Moreover, he conjectured that the theory
should exhibit a stringy behavior in this limit.  Now, this suggestion has been
made precise.  In principle, the complete $\lambda$ dependence of ${\mathcal N}
= 4$ gauge theory in the `t Hooft limit should be given by {\em classical} type
IIB superstring theory on $AdS_5 \times S^5$.  Many people are currently
studying this.

Finite temperature gauge theory is described by Euclideanizing the time
coordinate and taking it to be periodic.  Witten has shown that Euclideanized
AdS space, which has this structure on its boundary, contains a black hole at
the same temperature~\cite{witten98}.  I think that much more remains to be learned from
studying this correspondence.

An important concept that has emerged in recent years, called the {\em
holographic principle}~\cite{thooft93}, 
is incorporated by AdS/CFT duality.  This concerns the
number and location of degrees of freedom in a theory.  In a local quantum
field theory, the locality implies that
the number of degrees of freedom in a spatial region is
proportional to its volume.  However, this cannot be correct for a quantum
gravity theory, where the maximum entropy in a region is proportional to its
surface area.  (This bound is saturated in the case of a black hole.)  So the
idea of the holographic principle is that the physics in a region of space can
be encoded holographically on a surface that surrounds it.  This is what
happens in the case of AdS/CFT duality.  The physics of the AdS bulk 
(given by superstring theory) is not a
local QFT; rather, it is projected onto the boundary theory, which is a local
QFT.

The subject of AdS/CFT duality is still in its early days and developing
rapidly.  Whatever else I say is likely to be outdated by the time this is
published.  Suffice it to say that people are exploring all sorts of
generalizations.   These include breaking supersymmetries and conformal
symmetries and constructing analogous systems for $SO(N)$ and $Sp(N)$ gauge
theories.  So far, AdS/CFT duality has taught us more about gauge theories than
it has about string theory. This has the curious consequence that it may have
some useful spinoffs for the study of gauge theories such as QCD. This would
not require constructing a quantum vacuum that gives a realistic description
of all forces, only that it incorporates a reasonably good description of the gauge
theory one wants to study. This approach to the study of gauge theories
might turn out to be a useful alternative
to lattice gauge theory, though that
remains to be demonstrated.  I'm not sure whether it is more accurate 
to say that the second superstring revolution
is still going strong or that the third one has begun.

\section{Conclusion}

Beyond gauge theories there is superstring theory and M theory, 
but beyond superstring theory and M theory there are gauge theories.


\begin{thebibliography}{999}

\bibitem{jacob} {\em Dual Theory}, ed. M. Jacob, {\em Phys. Reports} reprint
volume (North-Holland 1974).

\bibitem{scherk75} J. Scherk, {\em Rev. Mod. Phys.} {\bf 47} (1975) 123.

\bibitem{yoneya74} T. Yoneya, {\em Prog. Theor. Phys.} {\bf 51} (1974) 1907.

\bibitem{scherk74} J. Scherk and J.H. Schwarz, {\em Nucl. Phys.} {\bf B81} (1974) 118.

\bibitem{candelas85} P. Candelas, G.T. Horowitz, A. Strominger, and E. Witten,
{\em Nucl. Phys.} {\bf B258} (1985) 46.

\bibitem{green87} M.B. Green, J.H. Schwarz, and E. Witten, {\em Superstring Theory},
2 vols., (Cambridge U. Press 1987).

\bibitem{hull94}  C. Hull and P. Townsend,
{\em Nucl. Phys.} {\bf B438} (1995) 109, hep-th/9410167.

\bibitem{witten95a} E. Witten,  {\em Nucl. Phys.} {\bf B443} (1995) 85, hep-th/9503124.

\bibitem{horava95} P. Ho\v{r}ava and E. Witten, {\em Nucl. Phys.} {\bf B460} (1996) 506,
 hep-th/9510209.

\bibitem{sen98} A. Sen, hep-th/9802051.

\bibitem{townsend95a} P.K. Townsend, {\em Phys. Lett.} {\bf B350} (1995) 184, hep-th/9501068.

\bibitem{cremmer78a} E. Cremmer, B. Julia, and J. Scherk, {\em Phys. Lett.} {\bf 76B}
(1978) 409.

\bibitem{giveon94}  For a review see 
A. Giveon, M. Porrati, and E. Rabinovici, {\em Phys. Rept.}
{\bf 244} (1994) 77, hep-th/9401139.

\bibitem{duff95}  For a review see 
M.J. Duff, R.R. Khuri, and J.X. Lu, {\em Phys. Rept.}
{\bf 259} (1995) 213, hep-th/9412184.

\bibitem{banks96} T. Banks, W. Fischler, S. Shenker, and L. Susskind, 
{\em Phys. Rev.} {\bf D55} (1997) 112, hep-th/9610043; 
For reviews see T. Banks,  hep-th/9710231; D. Bigatti and L. Susskind, hep-th/9712072.

\bibitem{susskind97} L. Susskind, hep-th/9704080.

\bibitem{polchinski95} J. Polchinski, {\em Phys. Rev. Lett.} 
{\bf 75} (1995) 4724,  hep-th/9510017;
p. 293 in {\em Fields, Strings, and Duality}
(TASI 96), eds. C. Efthimiou and B. Greene, World Scientific 1997, hep-th/9611050.

\bibitem{maldacena96} For reviews see J. Maldacena,  hep-th/9607235; 
D. Youm, hep-th/9710046; A. Peet, hep-th/9712253.

\bibitem{giveon98}  For a review see 
A. Giveon,  and D. Kutasov, {\em Nucl. Phys. Proc. Suppl.}
{\bf 68} (1998) 310, hep-th/9802067.

\bibitem{seiberg94} N. Seiberg and E. Witten, {\em Nucl. Phys. } 
{\bf B426} (1994) 19, hep-th/9407087;  {\em Nucl. Phys. } 
{\bf B431} (1994) 484, hep-th/9408099.

\bibitem{maldacena97} J.M. Maldacena,  hep-th/9711200.

\bibitem{klebanov97} I.R. Klebanov, {\em Nucl. Phys.} {\bf B496} (1997) 231, hep-th/9702076;
S.S. Gubser, I.R. Klebanov, and A.A. Tseytlin, {\em Nucl. Phys.} {\bf B499} (1997) 217,
hep-th/9703040; S.S. Gubser and I.R. Klebanov, {\em Phys. Lett.} 
{\bf B413} (1997) 41, hep-th/9708005; A.M. Polyakov, hep-th/9711002.

\bibitem{gubser98} S.S. Gubser, I.R. Klebanov,
and A.M. Polyakov, {\em Phys. Lett.} {\bf B428} (1998) 105,  hep-th/9802109;
E. Witten, hep-th/9802150.

\bibitem{freund} P. Freund and M. Rubin, {\em Phys. Lett.} {\bf 97B} (1980) 233;
for a review see M.J. Duff, B.E.W. Nilsson, and C.N. Pope,
{\em Phys. Rept.} {\bf 130} (1986) 1.

\bibitem{petervn} K. Pilch, P. van Nieuwenhuizen, and P.K. Townsend,
{\em Nucl. Phys.} {\bf B242} (1984) 377.

\bibitem{kim} H.J. Kim, L.J. Romans, and P. van Nieuwenhuizen,
{\em Phys. Rev.} {\bf D32} (1985) 389; M. G\"unaydin and N. Marcus, {\em Class. Quant. Grav.} {\bf 2} (1985) L11.

\bibitem{brink} L. Brink, J.H. Schwarz, and J. Scherk, {\em Nucl. Phys.} {\bf B121}
(1977) 77; F. Gliozzi, J. Scherk, and D. Olive, {\em Nucl. Phys.} {\bf B122} (1977) 253.

\bibitem{mandelstam} S. Mandelstam,  {\em Nucl. Phys.} {\bf B213} (1983) 149;
L. Brink, O. Lindgren, and B.E.W. Nilsson, {\em Phys. Lett.} {\bf 123B} (1983) 323;
P.S. Howe, K.S. Stelle, and P.C. West, {\em Phys. Lett.} {\bf 124B} (1983) 55;
P.S. Howe, K.S. Stelle, and P.K. Townsend, {\em Nucl. Phys.} {\bf B236} (1984) 125.

\bibitem{thooft74} G. 't Hooft, {\em Nucl. Phys.} {\bf B72} (1974) 461.

\bibitem{witten98} E. Witten, hep-th/9803131.

\bibitem{thooft93} G. 't Hooft, gr-qc/9310026; L. Susskind, {\em J. Math. Phys.} {\bf 36}
(1995) 6377, hep-th/9409089.

\end{thebibliography}
\end{document}